\PassOptionsToPackage{hyphens}{url}
\documentclass{llncs}
\usepackage[
  colorlinks=true,
  allcolors=black,
  pdftitle={BabelView: Evaluating the Impact of Code Injection Attacks in Mobile Webviews},
  pdfauthor={Claudio Rizzo, Lorenzo Cavallaro, Johannes Kinder},
  pdfsubject={Computer Science},
  pdfkeywords={Static analysis,Webview,Hybrid App,Android,Injection,Java,JavaScript,Interface,Bridge},
]{hyperref}

\usepackage{libertine}
\usepackage{inconsolata} 
\usepackage{libertinust1math}
\usepackage[T1]{fontenc} 
\usepackage{microtype}

\usepackage[all]{hypcap} 
\usepackage{graphicx}

\usepackage{algorithm}
\usepackage[noend]{algpseudocode}

\usepackage{booktabs} 

\def\Snospace~{\S{}}

\usepackage{cite}
\usepackage{tabularx}

\usepackage{tikz}
\usepackage{textcomp}
\usetikzlibrary{shapes,fit,calc,arrows.meta,positioning}

\usepackage{listings}
\usepackage{letltxmacro}
\newcommand*{\SavedLstInline}{}
\LetLtxMacro\SavedLstInline\lstinline
\DeclareRobustCommand*{\lstinline}{%
  \ifmmode
    \let\SavedBGroup\bgroup
    \def\bgroup{%
      \let\bgroup\SavedBGroup
      \hbox\bgroup
    }%
  \fi
  \SavedLstInline
}

\let\llncssubparagraph\subparagraph
\let\subparagraph\paragraph
\usepackage[compact]{titlesec}
\let\subparagraph\llncssubparagraph

\usepackage{siunitx}
\usepackage{standalone}

\newcommand{\n}[1]{\num{#1}}

\usepackage[nomessages]{fp}

\newcommand{\processed}{25000}

\newcommand{\totcrash}{1286}

\newcommand{\tofd}{3837}

\FPeval{\nocrash}{clip(\processed-\totcrash)}

\FPeval{\noto}{clip(\nocrash-\tofd)}

\newcommand{\nojs}{832}

\FPeval{\ifapps}{clip(\noto-\nojs)}

\newcommand{\totinstallations}{\n{3} billion}

\newcommand{\pos}{4997}
\FPeval{\nopos}{clip(\ifapps-\pos)}

\FPeval{\vulnrate}{round(100*(\pos/\ifapps), 1)}

\newcommand{\cfd}{810}

\newcommand{\caj}{0}

\FPeval{\crem}{clip(\totcrash-\cfd-\caj)}

\newcommand{\totvulns}{10808}

\newcommand{\validated}{50}
\newcommand{\TP}{42}
\newcommand{\FP}{10}
\newcommand{\TN}{1494}
\newcommand{\FN}{5}

\newcommand{\tp}{19}
\newcommand{\fp}{2}
\newcommand{\tn}{29}
\newcommand{\fn}{0}

\FPeval{\vprecision}{round((100*\TP/(\TP+\FP)), 0)}
\FPeval{\vrecall}{round((100*\TP/(\TP+\FN)), 0)}

\FPeval{\aprecision}{round((100*\tp/(\tp+\fp)), 0)}
\FPeval{\arecall}{round((100*\tp/(\tp+\fn)), 0)}

\newcommand{\mallo}{61.5}
\newcommand{\http}{98.7}
\newcommand{\jsinj}{482}
\newcommand{\jsnoiface}{1275}

\usepackage[absolute,showboxes]{textpos}

\TPMargin{5pt}

\begin{document}
    \begin{textblock}{0.566}(0.22,0.86)    
  \noindent
  \footnotesize\copyright{} Springer Verlag 2018. Published in Proceedings of
  the International Symposium on Research in Attacks, Intrusions, and Defenses
  (RAID 2018). This is the authors' version. The final authenticated publication
  is available online at \url{https://doi.org/10.1007/978-3-030-00470-5_2}.
    \end{textblock}

\title{BabelView: Evaluating the Impact of Code Injection Attacks in Mobile Webviews}

\author{Claudio Rizzo \and Lorenzo Cavallaro \and Johannes Kinder}
\institute{Royal Holloway, University of London\\
\email{\{claudio.rizzo.2015|lorenzo.cavallaro|johannes.kinder\}@rhul.ac.uk}
}
\maketitle

\begin{abstract}
  A Webview embeds a fully-fledged browser in a mobile application and allows that
  application to expose a custom interface to JavaScript code. This is a popular
  technique to build so-called hybrid applications, but it circumvents the usual
  security model of the browser: any malicious JavaScript code injected into the
  Webview gains access to the custom interface and can use it to manipulate the device
  or exfiltrate sensitive data.
  In this paper, we present an approach to systematically evaluate the possible
  impact of code injection attacks against Webviews using static information
  flow analysis. Our key idea is that we can make reasoning about JavaScript
  semantics unnecessary by instrumenting the application with a model of
  possible attacker behavior---the BabelView.
  We evaluate our approach on \n{\processed{}} apps from various Android
  marketplaces, finding \n{\totvulns{}} potential vulnerabilities in \n{\pos{}} apps. Taken
  together, the apps reported as problematic have over \totinstallations{}
  installations worldwide. We manually validate a random sample of \n{\validated{}}
  apps and estimate that our fully automated analysis achieves a precision of
  \n{\vprecision{}}\% at a recall of \n{\vrecall{}}\%.

  \keywords{Webview \and JavaScript interface \and Injection \and Static analysis}
\end{abstract}

\section{Introduction}

The integration of web technologies in mobile applications enables rapid
cross-platform development and provides a uniform user experience across
devices. Web content is usually rendered by a \textit{Webview}, a user
interface component with an embedded browser engine (\lstinline+WebView+ in
Android, \lstinline+UIWebView+ in iOS). Webviews are widely used: in 2015, about
85\% of applications on Google's Play Store contained
one~\cite{mutchler15:mobilewebapps}. Cross-platform frameworks such as Apache
Cordova, which allow apps to be written entirely in HTML and JavaScript, have
contributed to this high rate of adoption and given rise to the notion of \textit{hybrid applications}. 
Even otherwise native applications often embed Webviews for
displaying login screens or additional web content.

Unfortunately, Webviews bring new security
threats~\cite{LuoHDWY11,LuoJAD12, Thomas15,mutchler15:mobilewebapps}. 
While the Android Webview uses WebKit to render the page,
the security model can be modified by app developers. Whereas standalone browsers
enforce strong isolation, Webviews can intentionally poke holes in the browser
sandbox to provide access to app- and device-specific features via a
\textit{JavaScript interface}. For instance, a hybrid banking application could
provide access to account details when loading the bank's website in a Webview,
or it could relay access to contacts to fill in payee details.

For assessing the overall security of an application, it is necessary to
understand the implications of its JavaScript interface. When designing the
interface, a developer thinks of the functionality required by her own, trusted
JavaScript code executing in the Webview. However, there are several ways that
an attacker can inject malicious JavaScript and access the
interface~\cite{FahlHMSBF12,mutchler15:mobilewebapps}.

The observation that exposed interfaces can pose a security risk was made in
previous work~\cite{HassanshahiJYSL15,Bhavani13-corr-xss};  
however, not all interfaces are dangerous or offer meaningful
control to an attacker. The intuition is that flagging up---or even removing
from the marketplace---any applications with an exposed JavaScript interface
would be an excessive measure. By assessing the risk posed by an application, we
can focus attention on the most dangerous cases and provide meaningful
feedback to developers.

We rely on static analysis to evaluate the potential impact of an attack
against Webviews, with respect to the nature of the JavaScript interfaces. Our
key idea is that we can instrument an application with a model of potential
attacker behavior that over-approximates the possible information flow
semantics of an attack.
In particular, we instrument the target app and replace Android's Webview and its descendants with a
specially crafted \emph{BabelView} that simulates arbitrary interactions with
the JavaScript interface. A subsequent information flow analysis on the instrumented
application then yields new flows made possible by the attacker model, which
gives an indication of the potential impact. Together with an evaluation of the
difficulty of mounting an attack, this can provide an indication of the
overall security risk. 

Instrumenting the target application allows us to build on existing 
mature tools for Android flow analysis. This design makes our approach
particularly robust, which is important on a quickly changing platform such as
Android. In addition, since our instrumentation is over-approximate, we inherit
any soundness guarantees offered by the flow analysis used.
Independently of us, Yang et al.~\cite{YangMZG17} developed a
related approach to address the same problem, but with a closed source system
relying on a custom static analysis.
Our paper makes the following contributions:
\begin{itemize}
\item We introduce BabelView, a novel approach to evaluate
  the impact of code injection attacks against Webviews based on information
  flow analysis of applications instrumented with an attacker model. 
  BabelView is implemented using Soot~\cite{Vallee-RaiCGHLS99} 
  and is available as open source.
\item We analyze \n{\processed{}} applications from the Google Play Store to
  evaluate our approach and survey the current state of Webview security in
  Android. Our analysis reports \n{\totvulns{}} potential vulnerabilities in
  \n{\pos{}} apps, which together are reported to have more than
  \totinstallations{} installations. We validate the results on a random sample
  of \n{\validated{}} applications and estimate the precision to be
  \n{\vprecision{}}\% with a recall of \n{\vrecall{}}\%, confirming
  the practical viability of our approach.
\end{itemize}

In the remainder of the paper, we briefly explain Android WebViews
(\autoref{sec:background}) and introduce our approach (\autoref{sec:overview})
before describing the details of our implementation
(\autoref{sec:babelview}). We evaluate BabelView and report the results of our
Android study (\autoref{sec:evaluation}) and discuss limitations
(\autoref{sec:limitations}). Finally, we present related work
(\autoref{sec:relatedwork}) and conclude (\autoref{sec:conclusion}).

\section{Android WebViews}
\label{sec:background}

To provide the necessary context for the remainder of the paper, we first
introduce key aspects of Android Webviews.
An Android application can instantiate a Webview by calling its constructor or by declaring it
in the Activity XML layout, from where the framework will create it
automatically. The specifics of how the app interacts with the Webview object
depend on which approach it follows; in either case, a developer can extend
Android's \lstinline+WebView+ class to override methods and customize its behavior.

The \lstinline+WebView+ class offers mechanisms for interaction between the app
and the web content in both directions.
Java code can execute arbitrary JavaScript code in the Webview by passing a URL
with the ``\verb+javascript:+'' pseudo-protocol to the \lstinline+loadUrl+ method of
a Webview instance.  Any code passed in this way is executed in the context of
the current page, just like if it were typed into a standalone browser's address
bar.
For the other direction, and to let JavaScript code in the Webview call Java
methods, the Webview allows to create custom interfaces.  Any methods of an
object (the \textit{interface object}) passed to the \lstinline+WebView.addJavascriptInterface+ method that are tagged with the
\lstinline+@JavascriptInterface+ annotation\footnote{The
  \lstinline+@JavascriptInterface+ annotation was introduced in API level 17 to
  address a security vulnerability that allowed attackers to execute arbitrary
  code via the Java reflection API~\cite{webview-rce}.}  (the
\textit{interface methods}) are exported to the global JavaScript namespace in
the Webview. For instance, the following example makes a single Java method
available to JavaScript:
\begin{lstlisting}[numbers=none, frame=none]
LocationUtils lUtils = new LocationUtils();
wView.addJavascriptInterface(lUtils, "JSlUtils");

public class LocationUtils {
  @JavascriptInterface
  public String getLocation() { 
    do_something(); 
  }
}
\end{lstlisting}
Here, \lstinline+LocationUtils+ is bound to a global JavaScript object
\lstinline+JSlUtils+ in the Webview \lstinline+wView+. JavaScript code can
access the annotated Java method \lstinline+getLocation()+ by calling
\lstinline+JSlUtils.getLocation()+.

The Webview's JavaScript interface mechanism enforces a policy of which
Java methods are available to call from the JavaScript context. Developers of
hybrid apps are left to decide which functionality to expose in an
interface that is more security-critical than it appears. It is easy for a
developer to erroneously assume the JavaScript interface to be a trusted internal
interface, shared only between the Java and JavaScript portions of
the same app. In reality, it is more akin to a public API, considering the
relative ease with which malicious JavaScript code can make its way into a
Webview (see~\autoref{sec:attacker_model}).
Therefore, care must be taken to restrict the interface as much as possible and
to secure the delivery of web content into the Webview. In this work we provide a
way for developers and app store maintainers to detect applications with
insecure interfaces susceptible to abuse; our study in \autoref{sec:evaluation}
confirms that this is a widespread phenomenon.

\section{Overview}
\label{sec:overview}

We now introduce our approach by laying out the attacker model
(\autoref{sec:attacker_model}), describing our instrumentation-based model for
information flow analysis (\autoref{sec:instrumentation_overview}), and
discussing how the instrumentation preserves the application semantics
(\autoref{sec:preserving_semantics}).

\subsection{Attacker Model}
\label{sec:attacker_model}
Our overall goal is to identify high-impact vulnerabilities in Android
applications. Our insight is that injection vulnerabilities are difficult to
avoid with current mainstream web technologies, and that their presence does not
justify blocking an app from being distributed to users. Indeed, any standalone
browser that allows loading content via insecure HTTP has this vulnerability
(while calling this a ``vulnerability'' may be controversial, it clearly has
security implications and has led to an increasing adoption of HTTPS by
default). The ubiquity of advertisement libraries in Android apps further increases 
the likelihood of foreign JavaScript code gaining access to JavaScript interfaces.
Following this insight, we aim to pinpoint the risk of using a Webview that is
embedded in an app. To do this, we assess the
\textit{degrees of freedom} an attacker gains from injecting code into a Webview
with a JavaScript interface, which determines the potential impact of an
injection attack.

Consequently, the attacker model for our analysis consists of arbitrary code
injection into the HTML page or referenced scripts loaded in the Webview. In our
evaluation, we actively try to inject JavaScript into the Webview---e.g., as man
in the middle (see~\autoref{sec:feasibility}). We note, however, that other
channels are available to manipulate the code loaded into a Webview, including
malicious advertisements or site-specific cross-site-scripting
attacks~\cite{Bhavani13-corr-xss,HassanshahiJYSL15,JinHYDYP14}.
To abuse the JavaScript interface, the attacker then only requires the names of
the interface methods, which can be obtained through reverse-engineering.
Note that even a man in the middle becomes more powerful with access to the
JavaScript interface: the interface can allow the attacker to manipulate and retrieve
application and device data that would not normally be visible to the adversary.
For instance, consider a remote access application
with an interface method \lstinline+getProperty(key)+, which retrieves the value
mapped to a key in the application's properties. Without accessing the interface, 
an attacker may only ever observe calls to \lstinline+getProperty+ with, say, the keys \lstinline+"favorites"+ and
\lstinline+"compression"+, but the attacker would be free to also use the
function to retrieve the value for the key \lstinline+"privateKey"+.

\subsection{Instrumenting for Information Flow}
\label{sec:instrumentation_overview}
Our approach is based on static information flow (or taint) analysis. We aim to
find potentially dangerous information flows from injected JavaScript into
sensitive parts of the Java-based app and vice-versa.
At first glance, this appears to require expensive cross-language static
analysis, as recently proposed for hybrid
apps~\cite{BruckerH16,LeeDR16}.
However, we can avoid analyzing JavaScript code because our attacker model
assumes that the JavaScript code is controlled by the attacker. Therefore, we
want to model the actions performed by \textit{any possible JavaScript code},
and not that of developer-provided code that is supposed to execute in the
Webview.

To this end, we perform information flow analysis on the application
instrumented with a representation of the attacker model in Java, such that the
result is an over-approximation of all possible actions of the attacker
(we discuss alternative solutions in~\autoref{sec:limitations}).
We replace the Android \lstinline+WebView+ class (and custom subclasses) with a
\textit{BabelView}, a Webview that simulates an attacker specific to the app's
JavaScript interfaces. We then apply a flow-, \mbox{field-,} and
object-sensitive taint analysis~\cite{ArztRFBBKTOM14} to detect
information flows that read or write potentially sensitive information as a
result of an injection attack.

\begin{algorithm}[t]
\caption{Information flow attacker model}\label{bvattackpseudo}
\begin{algorithmic}[1]
\Procedure{Attacker}{}
\While {true}
\State {\textbf{choose} iface $\in$ JS-interfaces}
\State {result $\leftarrow$ iface(\textit{source}(), \textit{source}(), \ldots)}
\State {\textit{sink}(result)}
\EndWhile
\EndProcedure
\end{algorithmic}
\end{algorithm}
The BabelView provides tainted input sources to all possible sequences of
interface methods and connects their return values to sinks, as shown in
\autoref{bvattackpseudo}. Here, \lstinline+source()+ and
\lstinline+sink()+ are stubs that refer to sources and sinks of the underlying
taint analysis.  
The non-deterministic enumeration of sequences of interface method invocations
is necessary since we employ a flow-sensitive taint analysis. This way, our
model also covers situations where the information flow depends on a specific
ordering of methods; for instance, consider the following example:

\begin{lstlisting}[numbers=none, frame=none]
  String id;

  @JavascriptInterface
  public void initialize() { 
    this.id = IMEI(); 
  }
  @JavascriptInterface 
  public String getId() { 
    return this.id; 
  }
\end{lstlisting}
Here, a call to \lstinline+initialize+ (line 4) must precede any invocation of
\lstinline+getId+ (line 8) to cause a leak of sensitive information (the
IMEI).
The flow-sensitive analysis correctly distinguishes different orders of
invocation, which helps to reduce false positives. In the
BabelView, the loop in~\autoref{bvattackpseudo} coupled with non-deterministic choice forces the analysis to
join abstract states and over-approximate the result of all possible invocation
orders.

\begin{figure}[t!]
  \centering
  \begin{tikzpicture}
    \tikzset{
      font={\fontsize{9pt}{12}\sffamily\selectfont}}
    \tikzstyle{line}=[draw]
    pics/vhsplit/.style n {
      code = {
        \node[text width=3cm, align=center] (A) at (0,0) {Android API};  
        \node[anchor=north west, text width=3cm, align=center] (B) at (A.south) {Sinks};
        \node[anchor=north east, text width=3cm, align=center] (C) at (A.south){Sources};
   		
        \draw (C.north west) -- (B.north east)
        	  (A.south) -- (B.south west);
        
       	\node[inner sep=0pt,draw,rounded corners,fit=(A)(B)(C)] {}; 
        
        \node[anchor=north, text width=3cm, align=center, minimum height=0.5cm] (D) at (B.south west) {};
        \node[draw, rounded corners, anchor=north, text width=3cm, align=center] (E) at (D.south) {JS Interfaces}; 
    	
    	\draw[->] ($(E.north)!0.5!(E.north east)$) -- ($(B.south west)!0.5!(B.south)$);
    	\draw[<-] ($(E.north)!0.5!(E.north west)$) -- ($(C.south east)!0.5!(C.south)$);
        \node[anchor=north, text width=3cm, align=center, minimum height=0.5cm] (D1) at (E.south) {};
        
        \node[draw, rounded corners, anchor=north, text width=3cm, align=center] (F) at (D1.south) {BabelView};
		\draw[<->] (F.north) -- (E.south);
        \node[anchor=north, text width=3cm, align=center, minimum height=0.5cm] (D2) at (F.south) {};
        \node[draw, rounded corners, anchor=north east, text width=3cm, align=center] (G) at (D2.south) {Sources};
        
        \node[draw, rounded corners, anchor=north west, text width=3cm, align=center] (H) at (D2.south) {Sinks};
        
        \draw[->] ($(F.south)!0.5!(F.south east)$) -- ($(H.north west)!0.5!(H.north)$);
    	\draw[<-] ($(F.south)!0.5!(F.south west)$) -- ($(G.north east)!0.5!(G.north)$);
    	
    	\draw[-, orange, densely dashed, very thick] (C.south) to[out=270,in=180] (D1.center); 
    	\draw[->, orange, densely dashed, very thick] (D1.center) to[out=0,in=90] (H.north);
    	
    	\draw[<-, cyan, very thick] (B.south) to[out=270,in=0] (D1.center);
    	\draw[-, cyan, very thick] (D1.center) to[out=180,in=90] (G.north);
     	
     	\node [align=right, anchor=north, text=cyan, rotate=90,outer sep=20pt] (L) at (D1.east) {Web $\rightarrow$ Device};
     	\node [align=right, anchor=south, text=orange, rotate=90, outer sep=20pt] (L) at (D1.west) {Device $\rightarrow$ Web};
      }
    }
  \end{tikzpicture}
  \caption{BabelView models flows between the attacker and sensitive sources and
    sinks in the Android API that cross the JavaScript interface.}
  \label{fig:approach}
\end{figure}
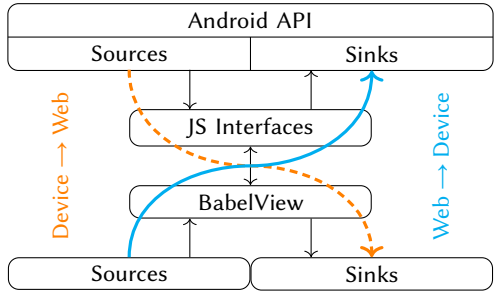

\autoref{fig:approach} illustrates our approach. We annotate certain methods in
the Android API as sources and sinks (see \autoref{sec:TaintAnalysis}), which
may be accessed by methods in the JavaScript interface. The BabelView includes
both a source passing data into the interface methods and a sink receiving their
return values to allow detecting flows both from and to JavaScript. The source
corresponds to any data injected by the attacker, and the sink to any method an
attacker could use to exfiltrate information, e.g., a simple web request.

\subsection{Preserving Semantics}
\label{sec:preserving_semantics}
Our instrumentation eliminates the requirement to perform a cross-language
taint analysis and moves all reasoning into the Java domain. However, we must
make sure that, apart from the attacker model, the instrumentation preserves
the original application's information flow semantics.
In particular, we need to integrate the execution of the attacker model into the
model of Android's application life cycle used as the basis of the taint
analysis~\cite{ArztRFBBKTOM14}. 
We solve this by overriding the methods used to load web content into the
Webview (such as \lstinline+loadUrl()+ and \lstinline+loadData()+) and modifying
them to also call our attacker model (\autoref{bvattackpseudo}). This is the
earliest point at which the Webview can schedule the execution of any injected
JavaScript code. The BabelView thus acts as a proxy simulating the effects of 
malicious JavaScript injected into loaded web content.

As the BabelView interacts only with the JavaScript interface methods, it does
not affect the application's static information flow semantics in any other way
than an actual JavaScript injection would. Obviously, this is not necessarily
true for other semantics: for example, the instrumented application would likely
crash if it were executed on an emulator or real device.

\section{BabelView}
\label{sec:babelview}

\begin{figure}[t]
  \centering
  \includegraphics[width=\textwidth]{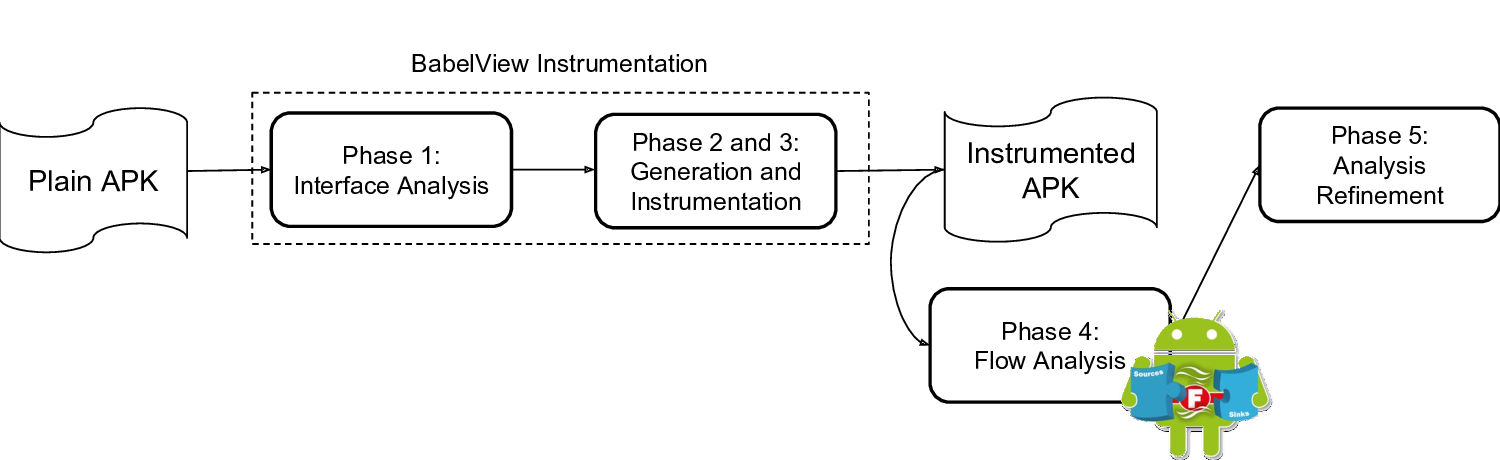}
  \vspace*{-24pt}
  \caption{Phases of our analysis.}
  \label{fig:phases}
\end{figure}

In this section, we explain the different phases of our
analysis. \autoref{fig:phases} provides a high-level overview: in Phase 1
(\autoref{sec:parsing}), we perform a static analysis to retrieve all interface
objects and methods, and associate them to the respective Webviews. In Phase 2
(\autoref{sec:generation}), we generate the BabelView, and, in Phase 3
(\autoref{sec:instrumentation}), we instrument the target application with
it. In Phase 4 (\autoref{sec:TaintAnalysis}), we run the taint flow analysis on
the resulting applications and finally, in Phase 5 (\autoref{sec:refinement}),
we analyze the results for flows involving the BabelView.

We implemented our static analysis and instrumentation using the Soot
framework~\cite{Vallee-RaiCGHLS99}; our information flow
analysis relies on FlowDroid~\cite{ArztRFBBKTOM14}. Overall, our
system adds about \n{6000} LoC to both platforms.

\subsection{Phase 1: Interface Analysis}
\label{sec:parsing}
As the first step of our analysis, we statically analyze the target application
to gather information about its Webviews and JavaScript interfaces. The goal of
this stage is to compute a mapping from Webview classes to classes of interface
objects that may be added to them at any point during execution of the app.

Using Soot, we can generate the application call graph and precisely resolve
callers and callees. We iterate through all classes and methods,
identifying all calls to \lstinline+addJavascriptInterface+, from where we then
extract Webviews that will hold interface objects. We make sure to treat
inheritance and polymorphism soundly in this stage, e.g., where parent classes
are used in variable declarations. We illustrate our approach to handle this on
the following code example:

\begin{lstlisting}[numbers=none, frame=none]
class FrameworkBridge {
  @JavascriptInterface
  public int foo() {...}
}

class MyBridge extends FrameworkBridge {
  @JavascriptInterface
  public int bar() {...}
}

class MyWebView extends WebView {...}

void initInterface(WebView aWebView, FrameworkBridge aBridge) { 
  aWebView.addJavascriptInterface(aBridge, "Android");
}

...
MyWebView mWebView = new MyWebView();
initInterface(mWebView, new MyBridge());
...
\end{lstlisting}
The code is adding the interface \lstinline+MyBridge+ to \lstinline+mWebView+,
an instance of \lstinline+MyWebView+.
The method \lstinline+initInterface+ is a wrapper (say, from a hybrid app
framework) that contains the actual call to \lstinline+addJavascriptInterface+.
When processing the call, we extract the types of \lstinline+aWebView+ and
\lstinline+aBridge+ from their parameter declarations.
For the Webview, we must process all descendants of its declared class to
include the types of all possible instances. For \lstinline+aWebView+, this
means we must instrument all descendants (including anonymous classes) of
\lstinline+WebView+, i.e., \lstinline+WebView+ and \lstinline+MyWebView+.

Similarly, we are interested in the type of \lstinline+aBridge+. Again, we must
iterate over all subclasses of its declared type \lstinline+FrameworkBridge+ to
ensure capturing the bridge added at runtime.
However, since \lstinline+addJavascriptInterface+ is of the unconstrained type
\lstinline+Object+, this could potentially include all classes. Therefore, we
restrict processing to just those subclasses that contain at least one
\lstinline+@JavascriptInterface+ annotation.
As a result, we obtain a superset of all interface objects that can be added by this
method, i.e., \lstinline+FrameworkBridge+ and \lstinline+MyBridge+.

Continuing the example, we now have the mapping from Webview classes to classes
of interface objects as
$\{
  \lstinline+WebView+ \mapsto \{ \lstinline+FrameworkBridge+, \lstinline+MyBridge+ \},
  \lstinline+MyWebView+ \mapsto \lstinline+FrameworkBridge+, \lstinline+MyBridge+ \}
\}$. Any additional occurrences of \lstinline+addJavascriptInterface+ will 
be processed analogously and the results added to the set.
Because the analysis in this phase is conservative in collecting compatible
types, the result is a sound over-approximation of the mapping of Webviews to
JavaScript interfaces that can occur at runtime (modulo inaccuracies from
dynamic code, see \autoref{sec:limitations}).

\subsection{Phase 2: Generating the BabelView}
\label{sec:generation}
We generate a \lstinline+BabelView+ class for each \lstinline+WebView+ in the
mapping. Each \lstinline+BabelView+ defines a subclass of its \lstinline+WebView+ (we
remove the parent's \lstinline+final+ modifier if necessary) and overrides all
of its parent's constructors so it can be used as a drop-in replacement.
We make the associated interface objects explicitly available in each
\lstinline+BabelView+. To this end, we override the
\lstinline+addJavascriptInterface+ method to store the interface objects passed
to it in instance fields of the \lstinline+BabelView+ class.

To implement the attacker model, the \lstinline+BabelView+ needs to
override all methods that load external resources and could thus be
susceptible to JavaScript injection. In particular, we override
\lstinline+loadUrl+, \lstinline+postUrl+, \lstinline+loadData+, and
\lstinline+loadDataWithBaseURL+. We automatically generate these 
methods as a call to their \lstinline+super+ implementation followed 
by a Java implementation of the attacker model,~\autoref{bvattackpseudo}. Finally,
the \lstinline+BabelView+ is equipped with two stub methods, \lstinline+leak+ and
\lstinline+taintSource+, representing a tainted sink and a tainted input,
respectively.  

\subsection{Phase 3: Instrumentation}
\label{sec:instrumentation}

In the next phase, we instrument the application to replace its Webviews with
our generated BabelView instances.
The instrumentation is case-dependent on how the Webview is instantiated
(see~\autoref{sec:background}): if it is created via an ordinary constructor
call, that constructor is replaced with the corresponding constructor of its
\lstinline+BabelView+ class.
If the Webview is created via the Activity XML layout, our instrumentation
searches for calls to \lstinline+findViewById+, which the app uses to
obtain the Webview instance (e.g., in order to add the JavaScript interface to
it).
To identify the calls to \lstinline+findViewById+ returning a Webview, our
instrumenter identifies explicit casts to a Webview class. Because we do not
parse the XML layout itself, we arbitrarily choose one of the constructors of
the \lstinline+BabelView+. While this
could potentially be a source of false positives or negatives, it would require
a highly specific and unconventional design of the Webview class.
\subsection{Phase 4: Information Flow Analysis}
\label{sec:TaintAnalysis}

We perform a static information flow analysis on the instrumented application to
identify information flows involving the attacker model. Since our approach
relies on instrumenting the application under analysis, it is agnostic to the
specific flow analysis.  We decided to rely on the open source implementation of
FlowDroid~\cite{ArztRFBBKTOM14}, inheriting its context-, flow-,
field-, and object-sensitivity, as well as its life cycle-awareness.

Sources and sinks are selected corresponding to sensitive information sources
and device functions, modified from the set provided by
SuSi~\cite{RasthoferAB14}. We further include the sources and
sinks used in the BabelView classes.

The information flow analysis abstracts the semantics of Android framework
methods. FlowDroid uses a simple modeling system (the \textit{TaintWrapper}),
where any method can either (i) be a source, (ii) be a sink, (iii) taint its
object if any argument is tainted and return a tainted value if its object is
tainted, (iv) clear taint from its object, (v) ignore any taint in its arguments
or its object.
We extended the TaintWrapper with several models that were relevant for the
types of vulnerabilities we were interested in, e.g., to precisely capture the
creation of Intents from tainted URIs.

Finally, information flows indicating that sensitive functionality is exposed
via the JavaScript interface are identified, triggering an alarm showing a
potential vulnerability. For instance, consider an \lstinline+Intent+ object
initialized to perform phone calls. A flow from \lstinline+source+ to
\lstinline+putExtra+ will taint the \lstinline+Intent+; if it is then passed as
an input to \lstinline+startActivity+, an attacker can perform calls on behalf
of the user.

\subsection{Phase 5: Analysis Refinement}
\label{sec:refinement}

\subsubsection*{Preferences.}
\label{sec:preferences}{}
Taint analysis cannot distinguish between individual
key-value pairs in a map.
\lstinline+Preferences+ are a commonly used map type in
Android apps that often store sensitive information as a key-value pair. After the information flow
analysis, we refine our results by statically deriving values of keys for access
to preferences.
Our definition of sources and sinks allows to identify both flows from and to
the \lstinline+Preferences+. Given two flows, one inserting and the other
retrieving values from \lstinline+Preferences+, we are interested in understanding
whether (i) the value is of the same type and (ii) the access key is the
same. If these conditions are met, we have identified a potential leak via
Preferences.
To determine the key values, we modeled \lstinline+StringBuilder+ and implemented
an intra-procedural constant propagation and folding for strings.
Finally, if an interface method allows web content to interact with a preferences object,
BabelView reports all keys used to access it, since preferences can be used to
store sensitive values.  This allows to inspect flows to or from preferences
entries, even if these values are not dependent on a specific source in the
Android API. We match key names against a list of suspicious entries, which can
highlight potential leaks of sensitive app-specific information
(see~\autoref{sec:eval_highlights}).
In the same manner, we also highlight suspiciously named interface methods.

\subsubsection*{Intents.}
\label{sec:intentAnalysis}

Flow analysis can detect situations where Intent creation depends on tainted
input.  However, it tells nothing about the type of the Intent created, as this
depends on specific parameters, e.g, those provided to its \lstinline+setAction+
method.  For interpreting results, it is important, however, to know the action
of an \lstinline+Intent+ that can be controlled by an attacker.
For any flow that reaches the \lstinline+startActivity+ sink, we perform an
inter-procedural backward dependency analysis to the point of the initialization
of the \lstinline+Intent+. If the \lstinline+Intent+ action is not set within the
constructor, we perform a forward analysis from the constructor to find calls to
\lstinline+setAction+ on the \lstinline+Intent+ object. The analysis may fail where
actions are defined within intent filters (XML definitions) or through other
built-in methods.
To increase precision in our inter-procedural analysis, we ensure that the
call-stack is consistent with an invocation through the interface method; i.e.,
the interface method that triggered the flow must be reachable.

\section{Evaluation}
\label{sec:evaluation}

We now present our evaluation of BabelView and the results of our study of
vulnerabilities in Android applications. Below, we explain our methodology
(\autoref{sec:eval_methodology}) and ask the following research questions to evaluate our
approach:
\begin{enumerate}
\item \textbf{Can BabelView successfully process real-world applications?} We
  conduct a study on a randomly selected set of applications from the AndroZoo~\cite{AllixBKT16}
  dataset and provide a breakdown of all results (\autoref{sec:eval_applicability}).
\item \textbf{Does BabelView expose real vulnerabilities?} We discuss some of
  the vulnerable apps in more detail to understand what an attacker can achieve
  under what conditions~(\autoref{sec:eval_highlights}).
\item \textbf{What are the precision and recall of our analysis?} We manually
  validate a random sample of apps, estimating overall precision and recall
  (\autoref{sec:eval_validation}).
\end{enumerate}
We also shed light on the current state of Webview security on Android with the
following questions:
\begin{enumerate}
\addtocounter{enumi}{3}
\item \textbf{How frequent are different types of alarms?} We report results per
  alarm, which provides an insight into the prevalence of potential
  vulnerabilities (\autoref{sec:eval_vulns}).
\item \textbf{Are there types of potential vulnerabilities that are likely to
    occur in combination?} We compute the correlation between alarms raised by
  our analysis and analyze our findings (\autoref{sec:eval_correlation}).
\end{enumerate}
Unfortunately, we were unable to conduct a direct comparison with BridgeScope,
the work most closely related to ours. Despite helpful communication, the
authors were ultimately unable to share neither their experimental data nor
their implementation with us. In the spirit of open data, we make all our code
and data available\footnote{\url{https://github.com/ClaudioRizzo/BabelView}}.

\subsection{Methodology} 
\label{sec:eval_methodology}

We obtained our dataset from AndroZoo~\cite{AllixBKT16}, using the list of applications
available on July 22nd, 2016, when it contained about 4.4 million samples.
We  downloaded a random subset of \n{209069}{}
apps, and then filtered our dataset for applications containing a Webview, a
call to \lstinline+addJavascriptInterface+, and granting permission to access
the Internet. As a result, we obtained \n{62674}{} total applications. Finally,
from the obtained sample, we randomly extracted \n{\processed{}}{} applications 
found in the Google Play Store, which we used for our analysis.

We ran BabelView on five servers: one 32-core with 250GiB of RAM and four
16-core with 125GiB of RAM. Each application took on average 180 seconds to
complete.  The high precision of FlowDroid's information flow analysis can lead
to long processing time in the order of hours. Therefore, we set a time limit of
15 minutes, which was a sweet spot in the sense that apps taking longer would
often go over an hour. A positive effect of our instrumentation-based approach
is that we benefit from improvements in the underlining flow analysis.  Indeed,
over the duration of this project, we saw a noticeable accuracy enhancement from
the constant improvements on FlowDroid.

Each application underwent three main phases: (i) BabelView instrumentation,
(ii) FlowDroid analysis on the instrumented app and (iii) analysis of the
resulting flows to identify suspect flows and raise alarms. On the reported
applications, we performed a feasibility analysis. We searched the app for plain
\verb+http://+ URLs and assess the resilience of the app against injection
attacks.

\subsection{Applicability}
\label{sec:eval_applicability}

Running our tool chain on the \n{\processed} target applications resulted
in \n{\totcrash{}} general errors and \n{\tofd} flow analysis timeouts. The remaining
\n{\noto{}} apps were successfully analyzed and we obtained the following
breakdown: \n{\nojs} applications had no interface objects at all or no
interface methods in case the target API was version 17 or above; \n{\nopos}
applications had no flows involving our attacker model; and \n{\pos} were
reported as dangerous, i.e., containing flows due to the attacker behavior.
This amounts to a rate of \n{\vulnrate}\%.
We investigated the reasons for the crashes, and most happened either due to
unexpected byte code that Soot fails to handle or while FlowDroid's taint
analysis was computing callbacks.

Among applications with interface objects, we also considered those targeting
outdated versions of the Android API, since this is still a common
occurrence~\cite{MutchlerSDM16,WuLXLG17,ThomasBR15}.
When using Webviews prior to API 17, any app is trivially vulnerable to an
arbitrary code execution disclosed in
2013\footnote{https://labs.mwrinfosecurity.com/blog/webview-addjavascriptinterface-remote-code-execution/}.
Despite targeting an old API version, if compiled with a newer Android SDK,
these applications can still use the \lstinline+@JavascriptInterface+
annotation. While the annotation itself does not provide extra security, these
apps may target newer APIs in future releases~\cite{Thomas15}.

\subsection{Alarms Triggered}
\label{sec:eval_vulns}

We successfully used BabelView to examine \n{\noto}{} applications. We found
that \n{\pos}{} of them triggered an alarm (i.e., our analysis reported a
potential vulnerability), meaning that the interface methods could be
exploited by foreign JavaScript from injection or advertisement. \autoref{tab:alarms}
shows a breakdown of all the alarms we observed in our analysis. Among the
most common alarms, we observed the possibility of writing to the File System
(Write File), capability to start new applications (Start
App), violation of the Same Origin Policy (Frame Confusion) and
the possibility of exploiting the old reflection attack due to Android API prior to v17.

\begin{table}[t]
\begin{tabularx}{\linewidth}{XS|XS|XS}
  \toprule
  \textbf{Alarm} & \textbf{\#Apps} & \textbf{Alarm} & \textbf{\#Apps} & \textbf{Alarm} & \textbf{\#Apps}\\
  \midrule
  Open File           &   385 & Write File        &  1444 & Read File            &   593\\
  TM Leaks            &    39 & Pref. TM Leaks    &     4 & Pref. Connectivity Leaks\hspace*{-3pt} &     4\\
  SQL-lite Leaks      &   136 & SQL-lite Query    &   438 & Pref. SQL-lite Leaks &    11\\
  GPS Leaks           &    43 & Pref. GPS Leaks   &     1 & Directly Send SMS    &     6\\
  Directly Make Calls &    19 & Call via Intent   &   314 & Email/SMS via Intent &   778\\
  Take Picture        &     7 & Download Photo    &   317 & Play Video/Audio     &   378\\
  Edit Calendar       &   357 & Post to Social    &   293 & Start App            &  1321\\
  API prior to 17     &  1039 & Unknown Intent    &  1107 & Frame Confusion      &  1039\\
  Fetch Class         &    85 & Fetch Constructor &     0 & Constructor init     &    13\\
  Fetch Method        &    85 & Method Parameter  &   622 & &\\
  \bottomrule
\end{tabularx}
\caption{Number of applications per alarm category. Pref. stands for indirect
  leaks via a Preference object; TM stands for Telephony Manager.}
\label{tab:alarms}
\end{table}

Writing File capabilities show the developers' need for storing app-external data usually
coming from an app-dedicated server. We also observed that many
applications implement advertising libraries which need to open a new application,
usually Google Play Store, to allow the user to download or visualize some
information. Unfortunately, the package name of the application to open is
given as input to an interface method, enabling a possible attacker to control
which app to start. Same-Origin-Policy violations are also very common: this is the case when a
\lstinline+loadUrl+ is invoked with input from the interface methods, 
controlling what is loaded in to a frame. As described
by Luo et al.~\cite{LuoHDWY11}, JavaScript executing in an iframe runs in
the context of the main frame, violating the SOP.

Many applications still target an API version prior
to 17~\cite{MutchlerSDM16,WuLXLG17,ThomasBR15}, 
often due to backwards compatibility or simply due to
confusion in declaring the SDK version.
Other alarms involve the possibility to prompt the user with an email or a text
message to send, directly sending an SMS or performing a phone call; prompting
the user with the call dialer; posting content to social network; interacting
with the calendar by creating or editing an event; playing videos or audio;
leaking sensitive information like the device ID or phone numbers (i.e. TM
Leaks), GPS position, SQL information, etc.

Finally, we shed light on the possible use of Java Reflection inside
interface methods. Fetch Class, Fetch Constructor, Constructor init, Fetch
Method and Method Parameter are all signs that an attacker controls
input used to execute methods via Java reflection. Although these are rare
situations and often hard to exploit, they are extremely high reward for an attacker as they
can potentially allow to circumvent the \lstinline+@JavascriptInterface+ annotation, leading
to arbitrary code execution.  
We manually analyzed some applications presenting
these alarms and in some cases an attacker could take control of a method and
its parameters, leading to remote code execution.

\subsection{Manual Validation}
\label{sec:eval_validation}

We used manual validation to estimate the accuracy of our analysis. In
particular we sampled and manually analyzed (i.e., reversed and
decompiled)~\n{\validated}{} applications. We evaluated two aspects:
\begin{enumerate}
  \item How accurate is BabelView with respect to each individual alarm it raises?
  \item Does BabelView function as an effective alarm system for hybrid apps?
\end{enumerate}

We began checking all types of alarms for each app and we established whether an
alarm was correctly triggered or correctly not triggered. We observed \n{\TP{}}
TPs (True Positives), \n{\FP{}} FPs (False Positives), \n{\TN{}} TNs (True
Negatives) and \n{\FN{}} FNs (False Negatives). From this, we can compute a
precision of \n{\vprecision{}}\% and a recall of \n{\vrecall{}}\% for our
analysis.

The results obtained are in line with our expectations. Our instrumentation does
not alter the semantics of applications other than adding a model of attack behavior.
Therefore, our precision depends on the underlining flow analysis.
However, more false positives could be introduced due to the 
object-insensitivity of our instrumentation---i.e., we distinguish types but not
instances of Webviews. Similarly, a very low false negative
rate is common for data flow analysis; however, FNs are still
possible, mainly due to incomplete Android framework.

To evaluate BabelView on a per-app basis, we consider a true positive the case
where an app contains at least one potential vulnerability and at least one
alarm is raised. True negatives and false positives/negatives follow
accordingly. We observed \n{\tp{}} TPs, \n{\fp{}} FPs, \n{\tn{}} TNs, and
\n{\fn} FNs, which yields a precision of \n{\aprecision{}}\% and a recall of
\n{\arecall{}}\%.
This suggest that BabelView performs well as an alarm system for potentially
dangerous applications. Even if individual alarms can be false positives, the
correlation of dangerous interfaces appears to leads to highlighted apps being
problematic with high probability.
The false negatives that are present when taken per vulnerability disappear 
when analyzed on a per app basis.

\subsection{Feasibility Analysis}
\label{sec:feasibility}

To better understand the feasibility of exploiting potential vulnerabilities
highlighted by BabelView, we measured the difficulty of performing an injection
attack. To this end we use a three-step process: (i) we check the application for
TLS misuse using MalloDroid~\cite{FahlHMSBF12}; (ii) we
search for hard-coded URLs beginning with \verb+http://+, suggesting that web
content could be loaded via an insecure channel; and (iii) we actively injected
JavaScript code into Webviews.

MalloDroid reported \n{\mallo}{}\% of applications using TLS
insecurely and \n{\http}{}\% of apps were found hard-coding HTTP URLs.
In order to actively inject JavaScript, we stimulated each reported application
with 100
Monkey\footnote{\url{https://developer.android.com/studio/test/monkey.html}}
events and actively intercepted the connection (using
Bettercap\footnote{\url{https://www.bettercap.org}}), trying to execute a
JavaScript payload. Moreover, we set up our own certificate authority and also
tried SSL strip attacks.
The goal of the injection was to determine whether the reported interface
methods were present in the Webview. To this end, we generated JavaScript code
checking for the presence of the interface objects reported by the BabelView
analysis. We were able to inject JavaScript in \n{\jsnoiface} applications and in
\n{\jsinj} cases we confirmed the presence of the vulnerable interface object.

\subsection{Correlation of Alarms}
\label{sec:eval_correlation}

\begin{figure}[t]
  \centering
  \includegraphics[
  width=\linewidth,
  trim=38 10 55 10,
  clip,
    ]{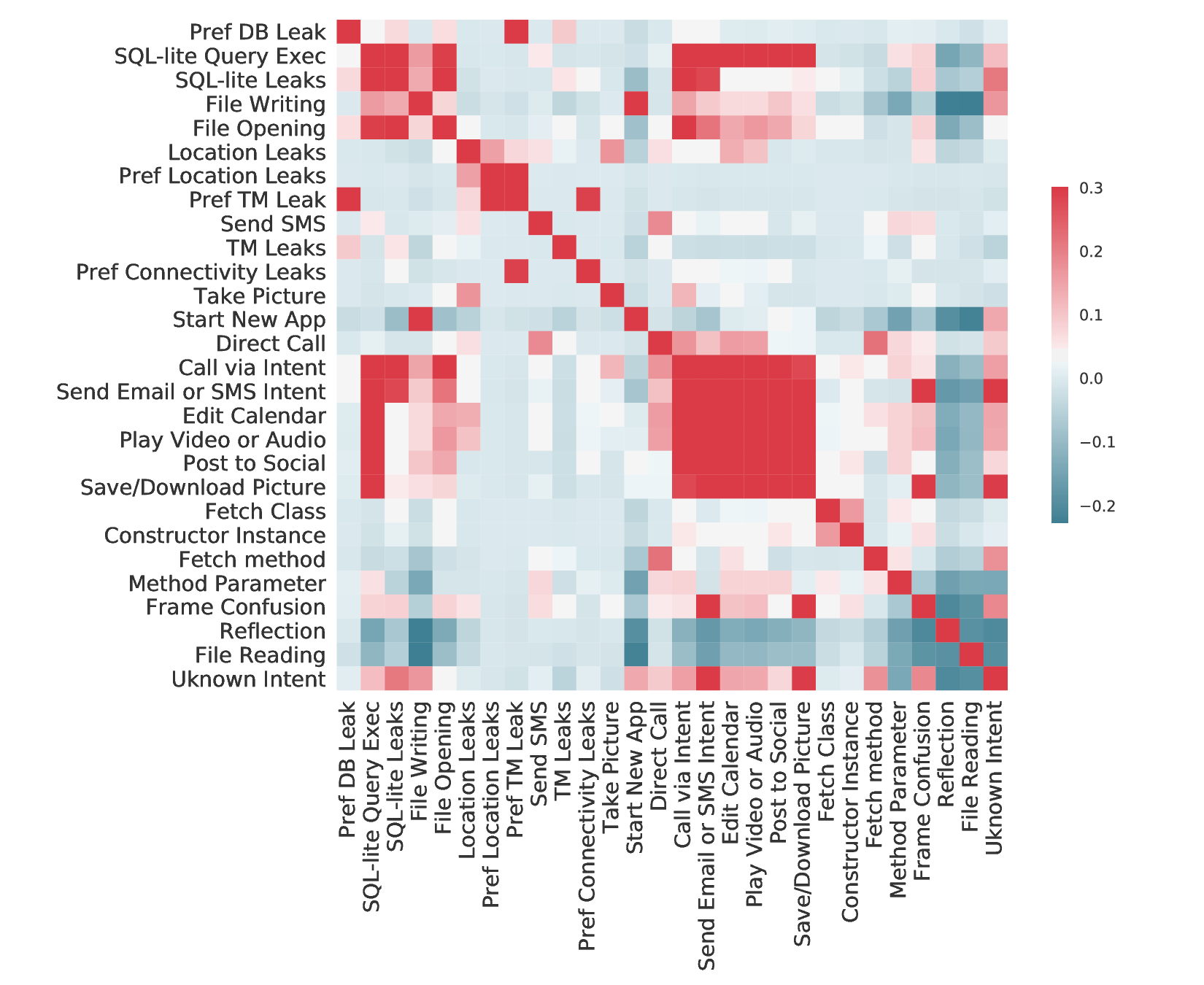}
  \caption{Correlation matrix of alarms.}
  \label{fig:heatmap}
\end{figure}

We were interested in finding correlations among the alarm categories we
identified. This does not only account for common patterns of functionality, but
also identifies single alarms that taken together could increase the attack
capabilities, e.g., combining opening and writing of a file results in writing
of arbitrary files.

We can see in the correlation matrix in~\autoref{fig:heatmap} that alarms
involving related functionality tend to be positively correlated (in red). For
example, opening and writing a file; SQL queries and leaks; and operations
involving intents such as call via intent, send email, edit calendar, play
video, post to social, and download pictures.  While some correlations are
evident, some appear incidental, such as intent calls and playing of
videos. Based on manual inspection (see ~\autoref{sec:eval_highlights}), we
found that these categories of alarms often appear together in apps using common
libraries, e.g., for advertisements.

\subsection{Individual Case Studies}
\label{sec:eval_highlights}
We now report individual case studies to illustrate the nature of our findings. 

\subsubsection*{Advertising Libraries.}

During the evaluation, we discovered an advertising library, used by \n{353} of
\n{\pos{}} applications, which implements a Webview exposing many sensitive
interface methods. In particular, a successful JavaScript injection would allow
an attacker to perform different actions, including downloading/saving of
pictures, sending email or SMS by manipulating \lstinline+Intents+, playing audio
or videos on the victim's phone, opening new applications, creating calendar
events, and posting to social networks.

Another library, used by \n{1507} applications, allows an attacker to
start new applications on the phone, controlling the \lstinline+Intent+ extras
provided to the \lstinline+Activity+. 

\subsubsection*{Game App.}

Among our results, we found a game application (``SwingAid Level up Golf'') that
uses several Webviews and JavaScript interfaces leading to different alarms:
SQL-lite leaks via preferences, frame confusion, and telephony manager
Leaks. Moreover, we discovered the value \emph{loginPwd} among preferences keys
accessible from a JavaScript interface. We were able to manually confirm all
alarms as true positives. Interface methods accessible when creating an account
creation within the game include \lstinline+getAccountEmail+,
\lstinline+getPhoneNumber+, and \lstinline+getUserPwd+.  We successfully
performed a man-in-the-middle attack and injected JavaScript to access all three
methods. The account e-email and phone number are accessible immediately upon
attempting to create an account. The password is stored in a local database,
cached in the preferences and accessible with the \emph{loginPwd} key. When the
user visits the account creation page a second time, the password can be stolen
via the interface method.

The underlying problem is twofold and representative for many Webview
vulnerabilities: first, the Webview loads data via an insecure channel, and
second, the JavaScript interface makes sensitive data available (a plaintext
password). Even if the password would otherwise not be sent via the insecure
channel, a JavaScript injection attack is able to retrieve it through the
interface and extract it directly.
Since our discovery, all issues have been resolved in a newer version of the
application (version 2.6).
  
\section{Limitations and Discussion}
\label{sec:limitations}

\subsubsection*{Avoiding Instrumentation}
In principle, we could avoid instrumenting the application by summarizing
interface methods with an interprocedural taint analysis.  However, to achieve
the same precision, the analysis would have to be computationally expensive: on
method entry, any reachable field in any reachable object (not just arguments of
the interface method) would have to be treated as carrying individual taint. On
method exit, the effects on all reachable fields would have to stored, before
resolving the effects among all interface method summaries.
Our instrumentation-based approach not only avoids this cost, but also allows us
to factor out flow analysis into a separate tool, a design choice that improves
robustness and maintainability.

\subsubsection*{Analysis Limitations}
Our system, together with the underlying flow analysis, is subject to common
limitations of static analysis and hence can fail to detect Webviews and
interfaces instantiated via native code, reflection, or dynamic code loading. In
principle, this currently allows a developer intent on doing so to hide
sensitive JavaScript APIs. However, we focus on benign software and
vulnerabilities that are honest mistakes rather than planted backdoors. Still,
we note that BabelView would automatically benefit from future flow
analyses that may counteract evasion techniques.

A potential source of false positives is that BabelView does not distinguish
Webview instances of the same type and will conservatively join the JavaScript
interfaces of all instances.
Furthermore, our analysis loses precision when reporting indirect leaks via
\lstinline+Preferences+ or \lstinline+Bundle+. As mentioned in
\autoref{sec:TaintAnalysis}, we connect sensitive flows into the application
preferences with flows from the preferences to the instrumented sink method in
BabelView. While this is sound and will conservatively capture any information
leaks via preferences, it is not taking into account any temporal dependencies
between storing and retrieving the value. A different treatment of this would be
a potential source of false negatives, since preferences persist across
application restarts.

\subsubsection*{Attack Feasibility} 

In our feasibility analysis, we actively try to inject JavaScript code into a
Webview, aiming at identifying whether the reported interface object is
present in the Webview. The presence of the interface object means that all
its interface methods are available to use, including the one BabelView
reported. However, we do not actively invoke these methods and thus we cannot
be sure of their exploitability.

\subsubsection*{Mitigating Potential Vulnerabilities}

To avoid giving potential attackers control over sensitive data and
functionality, developers can follow a set of design principles. First of all,
Webview contents should be exclusively loaded via a secure channel. Second, as
mentioned in the Android developer documentation, Webviews should only load
trusted contents. External links have to be opened with the default browser. For
also protecting against malicious ads or cross-site-scripting attacks,
JavaScript interfaces should offer an absolute minimum of functionality and
avoid arguments as far as possible. Finally, recent work also introduced novel
mechanisms to enforce policies on hybrid applications
(see~\autoref{sec:relatedwork/ac}).

\section{Related Work}
\label{sec:relatedwork}

We now review work on vulnerabilities and attacks against Webview
(\autoref{sec:relatedwork/attaks}), discuss related work on policies and access
control (\autoref{sec:relatedwork/ac}), and contrast with work on
instrumentation-based modeling (\autoref{sec:relatedwork/if}).

\subsection{Webview: Attacks and Vulnerabilities} 
\label{sec:relatedwork/attaks}
Webview vulnerabilities have been widely 
studied~\cite{LuoHDWY11,LuoJAD12,NeugschwandtnerLP13,ChinW13,mutchler15:mobilewebapps,Bhavani13-corr-xss}. 
Luo et al. give a detailed overview
of several classes of attacks against Webviews~\cite{LuoHDWY11}, 
providing a basis for our work. 
Neugschwandtner et al.~\cite{NeugschwandtnerLP13} were the first
to highlight the magnitude of the problem. In their analysis, they categorize as
vulnerable all applications implementing JavaScript interfaces and misusing TLS
(or not using it at all). For further precision, they analyzed permissions and
discovered that 76\% of vulnerable applications requested privacy critical
permissions. While this is a sign of poorly designed applications, the impact of
an injection exploit very much depends on the JavaScript interfaces, motivating
the work of this paper.

A step forward towards this was made by Bifocals~\cite{ChinW13},
a static analysis tool able to identify and evaluate vulnerabilities in
Webviews. Bifocals looks for potential Webview vulnerabilities (using
JavaScript interfaces and loading third party web pages) and then performs an
impact analysis on the JavaScript interfaces. In particular, it analyzes
whether these methods reach code requiring security-relevant
permissions. However, JavaScript interfaces can pose an
(application-specific) risk without making use of permissions. 
At the same time, not all JavaScript interfaces that make use of permissions are
dangerous: for example, an interface method might use the phone's IMEI to
perform an operation but not return it to the caller. 

The means by which malicious code can be injected into the Webview have been
discussed in previous
work~\cite{HassanshahiJYSL15,JinHYDYP14}.
Having to interact with many forms of entities, HTML5-based hybrid applications
expose a broader surface of attack, introducing new vectors of injection for
cross-site-scripting attacks~\cite{JinHYDYP14}. While these
attacks require the user to directly visit the malicious page within the
Webview, Web-to-Application injection attacks (W2AI) rely on intent
hyperlinks to render the payload simply by clinking a link in the default
browser~\cite{HassanshahiJYSL15}. Both discuss the threat
behind JavaScript interfaces, but stop their analysis at the moment where the
malicious payload is loaded, without analyzing the implication of the attacker
executing the JavaScript interfaces.

A large scale study on mobile web applications and their vulnerabilities was
presented by Mutchler et al.~\cite{mutchler15:mobilewebapps}, but did not study
the nature of the exposed JavaScript interfaces.
Li et al.~\cite{LiWZCWXBZH17} studied a new category of fishing
attacks called \emph{Cross-App WebView infection}. This new type
of attacks exploits the possibility of issuing navigation requests from one
app's Webview to another via Intent deep linking and other URL schemata. This
can trigger a chain of requests to a set of infected apps.

Most closely related to our work is the concurrently developed 
\emph{BridgeScope}~\cite{YangMZG17}, a tool to assess
JavaScript interfaces based on a custom static analysis. Similar to our work, BridgeScope
allows to detect potential flows to and from interface methods.  BridgeScope
uses a custom flow analysis, whereas our approach intentionally allows to reuse
state-of-art flow analysis tools. While BridgeScope's flow analysis performs
well on benchmarks, there appears to be no specific treatment of Map-like
objects such as \lstinline+Preferences+ of \lstinline+Bundle+.

In recent work, Yang et al.~\cite{EOEDroid_NDSS18} have combined the information
of a deep static analysis with a selective symbolic execution to actively
exploit event handlers in Android hybrid applications.  In
\emph{OSV-Hunter}~\cite{OSV_SP18}, they introduce a new approach to detect
Origin Stripping Vulnerabilities. These type of vulnerabilities persist
when upon invocation of \lstinline+window.postMessage+, it is not possible to
distinguish the identity of the message sender or even safely obtain the source
origin. This is inherently true for Hybrid applications, where developers often
rely on JavaScript interfaces to fill the gap between web and the native
platform.

\subsection{Webview Access Control}
\label{sec:relatedwork/ac}

There have been several proposals to bring origin-based access control to
Webviews~\cite{GeorgievJS14,TuncayDG16,ShehabJ14}. 
Shehab et al.~\cite{ShehabJ14} proposed a
framework that modifies Cordova, enabling developers to build and enforce a
page-based plugin access policy. In this way, depending on the page loaded, it
will or will not have the permission to use exposed Cordova plugins (i.e.,
JavaScript interfaces).

Georgiev et al. presented
NoFrank~\cite{GeorgievJS14}, a system to extend origin-based
access control to local resources outside the web browser. In particular, the
application developer whitelists origins that are then allowed to access
device's resources. However, once an origin is white-listed, it can access any
resource exposed. Jin et al.~\cite{JinWLD13} propose a fine-granular solution
in a system that allows developers to assign different
permissions to different frames in the Webview.

Tuncay et al.~\cite{TuncayDG16} increase granularity further in their Draco system. Draco
defines a policy language that developers can use to design access control
policies on different channels, i.e. the interface object, the event handlers
and the HTML5 API.
Another framework allowing developers to define security policies is
HybridGuard~\cite{PhungMRS17}. Differently from Draco, HybridGuard has been entirely
developed in JavaScript, making it platform independent and easy to deploy on
different platform and hybrid development framework.
Both Draco and HybridGuard could provide an interesting solution to the problem
of securing an interface BabelView is rising an alarm for, without unduly
restricting its functionality.

\subsection{Instrumentation-based Modeling}
\label{sec:relatedwork/if}

Synthesizing code to trigger specific function interfaces is not a new problem
and traces back to generating verification harnesses, e.g., for software model
checking~\cite{BallBCLLMORU06,bat-fmcad10}. On Android,
FlowDroid~\cite{ArztRFBBKTOM14} uses a model that invokes
callbacks in a ``dummy main'' method, taking into account the life cycle of
Android activities.
While the problems share some similarity, JavaScript interfaces and Webviews are
inherently varied and app-specific. Therefore, we require a static analysis and
cannot rely on fixed signatures.
Furthermore, because our model represents an attacker instead of a well-defined
system, calls can appear out of context anytime web content can be loaded in the
Webview, i.e., after a \lstinline+loadUrl+-like method.

\section{Conclusion}
\label{sec:conclusion}

In this paper, we presented a novel method to use information flow analysis to
evaluate the possible impact of code injection attacks against mobile
applications with Webviews. The key idea of our approach is to model the
possible effects of injected malicious JavaScript code at the Java level,
thereby avoiding any direct reasoning about JavaScript semantics. In particular,
this allowed us to rely on a robust state-of-the-art implementation of taint
analysis for Android.

We implemented our approach in BabelView, and evaluated it on \n{\processed{}}
applications, confirming its practical applicability and at the same time
reporting on the state of Webview security in Android. 
With BabelView, we found \n{\totvulns{}} potential vulnerabilities in \n{\pos{}}
applications, affecting more than \totinstallations{} users. We validated our
results on a subset of applications where we achieved a precision of
\n{\vprecision{}}\% at a recall of \n{\vrecall{}}\% when measured per alarm, or
a precision of \n{\aprecision{}}\% and a recall of \n{\arecall{}}\% when
measured per application.

\section*{Acknowledgments}
We would like to thank our shepherd, Angelos Stavrou, and the anonymous
reviewers for their valuable feedback. We are grateful to Roberto Jordaney,
Blake Loring, Duncan Mitchell, James Patrick-Evans, Feargus Pendlebury, and
Versha Prakash for their help and their comments on earlier drafts of this
paper.
This work was in part supported by EPSRC grant EP/L022710/1 and a Google Faculty
Award.

\bibliographystyle{abbrv}
\bibliography{bibliography}

\end{document}